\RequirePackage{lineno}
\documentclass[fleqn,aps,pre,preprint,a4paper,onecolumn,floatfix]{revtex4}

\usepackage{amsmath}
\usepackage{amssymb}

\usepackage[dvips]{graphicx}
\usepackage{color}
\usepackage{tabularx}
\usepackage{mathtools}
\usepackage{algpseudocode}
\usepackage{enumitem}
\usepackage{multirow}
\usepackage{epstopdf}  
\usepackage{bm}
\usepackage{url}
\usepackage[normalem]{ulem}

 \newcommand{\redc}[1]{{#1}}

\newcommand{\me}{\mathcal {E}}

\newcommand{\ma}{\mathcal {A}}

\usepackage{verbatim}

\begin{document}
\title{
  Efficient sampling of high-dimensional free energy landscapes using adaptive reinforced dynamics
  }
\author{Dongdong Wang}
\affiliation{Program in Applied and Computational Mathematics, 
Princeton University, Princeton, NJ 08544, USA}
\affiliation{DP Technology, Beijing, People’s Republic of China}
\author{Yanze Wang}
\affiliation{DP Technology, Beijing, People’s Republic of China}
\affiliation{College of Chemistry and Molecular Engineering, 
Peking University, Beijing, 100871, P.R.~China}
\author{Junhan Chang}
\affiliation{DP Technology, Beijing, People’s Republic of China}
\affiliation{College of Chemistry and Molecular Engineering, 
Peking University, Beijing, 100871, P.R.~China}
\author{Linfeng Zhang}
\email{linfeng.zhang.zlf@gmail.com}
\affiliation{Program in Applied and Computational Mathematics, 
Princeton University, Princeton, NJ 08544, USA}
\affiliation{DP Technology, Beijing, People’s Republic of China}
\author{Han Wang}
\email{wang_han@iapcm.ac.cn}
\affiliation{Laboratory of Computational Physics,
  Institute of Applied Physics and Computational Mathematics, Fenghao East Road 2, Beijing 100094, P.R.~China}
\author{Weinan E}
\affiliation{School of Mathematical Sciences, Peking University, Beijing, People’s Republic of China}
\affiliation{Department of Mathematics and Program 
in Applied and Computational Mathematics, 
Princeton University, Princeton, NJ 08544, USA}
\affiliation{Beijing Institute of Big Data Research, 
Beijing, 100871, P.R.~China}

\maketitle

\begin{quote}
\textsc{Abstract} 
Enhanced sampling methods such as metadynamics and umbrella sampling have become essential tools for exploring the configuration space of molecules and materials.  
At the same time, they have long faced a number of issues such as the inefficiency when dealing with a large number of 
collective variables (CVs) or systems with high free energy barriers. 
In this work, we show that with \redc{the clustering and adaptive tuning techniques}, the reinforced dynamics (RiD) scheme can be used to efficiently explore the configuration space and free energy landscapes with a large number of  CVs or systems with high free energy barriers.
We illustrate this by studying various representative and challenging examples.
Firstly we demonstrate the efficiency of adaptive RiD compared with other methods, and construct the 9-dimensional free energy landscape of peptoid trimer which has energy barriers of more than 8 kcal/mol.
We then study the folding of the protein chignolin using 18 CVs.
In this case, both the folding and unfolding rates are observed to be equal to 4.30~$\mu s^{-1}$.
Finally, we propose a protein structure refinement protocol based on RiD.
This protocol allows us to efficiently employ more than 100 CVs for exploring the landscape of protein structures and it gives rise to an overall improvement of 14.6 units over the initial Global Distance Test-High Accuracy (GDT-HA) score.
\end{quote}



\setlength{\parskip}{.5em}
\section{Introduction}
Over the past several decades, molecular dynamics (MD) has become an essential tool in modeling the structure and dynamics of biomolecules.
At the same time, it has also been recognized that an essential difficulty with MD is the time scale it can access, due to the presence of a large number of (free) energy barriers in the (free) energy landscape of these biomolecules, since crossing such barriers are rare events.
Enhanced sampling methods have thus been proposed to accelerate the sampling over the phase space. 
A useful idea has been to add a biasing potential, a function of one or more collective variables (CVs) of the system, to the potential energy, so that the free energy barriers are reduced. 
Well-known examples of the application of such ideas include metadynamics  (MetaD)~\cite{laio2002escaping,barducci2008well} and umbrella sampling~\cite{torrie1977nonphysical}.
Both of them have become popular tools in molecular simulations.
However, the effectiveness of these methods is much reduced as the number of CVs increases. 
When the number of CVs is small, the choice of the CVs becomes a critical issue for the accuracy of these methods.
Yet at the moment, there are still no systematic and reliable ways to choose such CVs.

An important advance was the temperature accelerated MD (TAMD)~\cite{rosso2002use,maragliano2006temperature,abrams2008efficient}. It proposes to treat the CVs as dynamical variables, couple them, adiabatically~\cite{rosso2002use} or harmonically~\cite{maragliano2006temperature,abrams2008efficient}, to the atomistic system by introducing an extended system, and accelerate the sampling by using an artificially higher temperature for these variables.
TAMD is a very effective tool for exploration. 
It allows for at least dozens of CVs to be used, and it has been successfully applied to large-scale systems~\cite{AbramsLargeSpaceTAMD}.
On the front of reconstructing the free energy surface, Maragliano et al~\cite{maragliano2008single} proposed to use radial-basis
functions, which predated the work using neural networks.
Whether these ideas can be used to accurately calculate the free energy of systems with a larger number of CVs as the ones reported here
is a non-trivial issue that remains to be addressed.

Recently, some efforts on dealing with larger number of CVs have been made under the MetaD framework.
Bias-exchange metadynamics (BEMetaD)~\cite{Bias-Exchange} is proposed to simulate multiple replicas at the same temperature, and each replica is biased with a different set of CVs.
Using the replica exchange idea, BEMetaD is able to construct free energy surfaces (FESs) in the high-dimensional CV space.
Parallel bias metadynamics (PBMetaD)~\cite{Pfaendtner2015Efficient,2018Biasing} constructs a high-dimensional bias potential by the weighted direct product of low-dimensional ones, and boosts the sampling with only one replica.
These methods can be implemented under the well-tempered MetaD scheme.
Both BEMetaD and PBMetaD assume that the biasing potential is a direct product, so they are effective only when the FES has a similar structure, which is not necessarily true for general high-dimensional problems.

Recent advances in machine learning (ML) offer a powerful tool for approximating high-dimensional functions. 
Several attempts have been made along this direction~\cite{stecher2014free,mones2016exploration,schneider2017stochastic,zhang2018reinforced,sidky2018learning,guo2018adaptive,sultan2018transferable,bonati2019neural,sevgen2020combined}.
These approaches represent the free energy surface by high-dimensional approximators like kernel functions or deep neural networks (DNNs).
What differentiates different approaches is
how one \emph{optimizes} the parameters of the approximators and how the approximators are used on-the-fly to facilitate  sampling.
The  Gaussian process regression is used to reconstruct the FES~\cite{stecher2014free} and to enhance the sampling of the system~\cite{mones2016exploration}. 
DNNs trained by stochastic optimization techniques are proposed to represent the FESs~\cite{schneider2017stochastic}.
The efficiency of sampling is enhanced by biasing the system with DNNs trained from density estimations~\cite{sidky2018learning} or from mean forces~\cite{guo2018adaptive}.
Sultan et.al.~proposed nonlinear latent embedding for obtaining the latent CVs~\cite{sultan2018transferable}.
The neural network-based variationaly enhanced sampling (VES) approach~\cite{bonati2019neural}  minimizes the Kullback-Leibler (KL) divergence between the distribution of the CVs and a target distribution~\cite{valsson2014variational,shaffer2016enhanced}.
The performance of these approximators has also been studied by Cendagorta et.~al.~\cite{cendagorta2020comparison}.
Their results show that the kernel method typically requires larger memory and its performance deteriorates as the number of  CVs
increases, while the neural network models are more robust and accurate.

The starting point of the present work is the  reinforced dynamics (RiD) scheme introduced in~\cite{zhang2018reinforced}.
RiD uses an uncertainty indicator to decide where to bias the MD simulation and at which values of the CVs one should calculate the mean force.
The accumulated CVs and mean forces are used for training and refining the DNN parameters.
RiD has been shown to be successful in exploring spaces with no more than 20 CVs, but its efficiency deteriorates quickly with more CVs: It has been observed getting trapped in the deep local minima for higher dimensional systems. 
This difficulty is caused by the following two factors: 
(1) The probability of visiting the neighborhood of a local minimum in a high-dimensional CV space is much lower than that in lower dimensional cases.
Thus a random batch of RiD samples is not enough for reconstructing the landscape near the local minimum of a high-dimensional FES.
(2) The biasing mechanism is too rigid for exploration, and particularly for  escaping from deep local minima.

Despite all these efforts, we are still unable to tackle challenging tasks such as protein structure refinement, in which MD simulations are used to refine the decoys predicted by data-driven methods.
Such a task usually requires repetitions of the unfolding and refolding processes, which takes a very long computational time because of the kinetic barriers.
Enhanced sampling methods such as replica exchange MD (REMD) simulation~\cite{sugita1999replica} and metadynamics could in principle be used to overcome the barriers.
In practice, REMD is very expensive and, for metadynamics, it is unclear how to choose  a low dimensional CV space even for small proteins with dozens of residues in the absence of the knowledge of the native structure. 
If one chooses  all the dihedral angles (twice the number of residues) as the CVs, then the approaches 
 discussed above may, in principle, be applied. 
However, in practice, to the best of our knowledge, these methods, have never been reported to be able to handle problems with more than 100 CVs, so their effectiveness in solving the protein structure refinement problem is still unknown.



In this work, we propose an adaptive version of the RiD scheme, which has the potential to overcome the difficulties with RiD,
 and we demonstrate its effectiveness on several challenging examples.
The novelty of the adaptive RiD lies in two aspects: (1) A clustering algorithm is applied to the configurations selected for labeling to reduce the number of configurations needed to 
represent the unexplored configuration space.
(2) The uncertainty indicator and the bias are adaptively and automatically tuned by using the same clustering argument that quantifies the diversity of the explored configurations.
This helps to encourage RiD to escape from the deep local minima.
In addition, adaptive RiD supports a multi-walker scheme, which allows to make full use of parallel computational platforms.



\begin{figure}
\centering
\includegraphics[width=.95\linewidth]{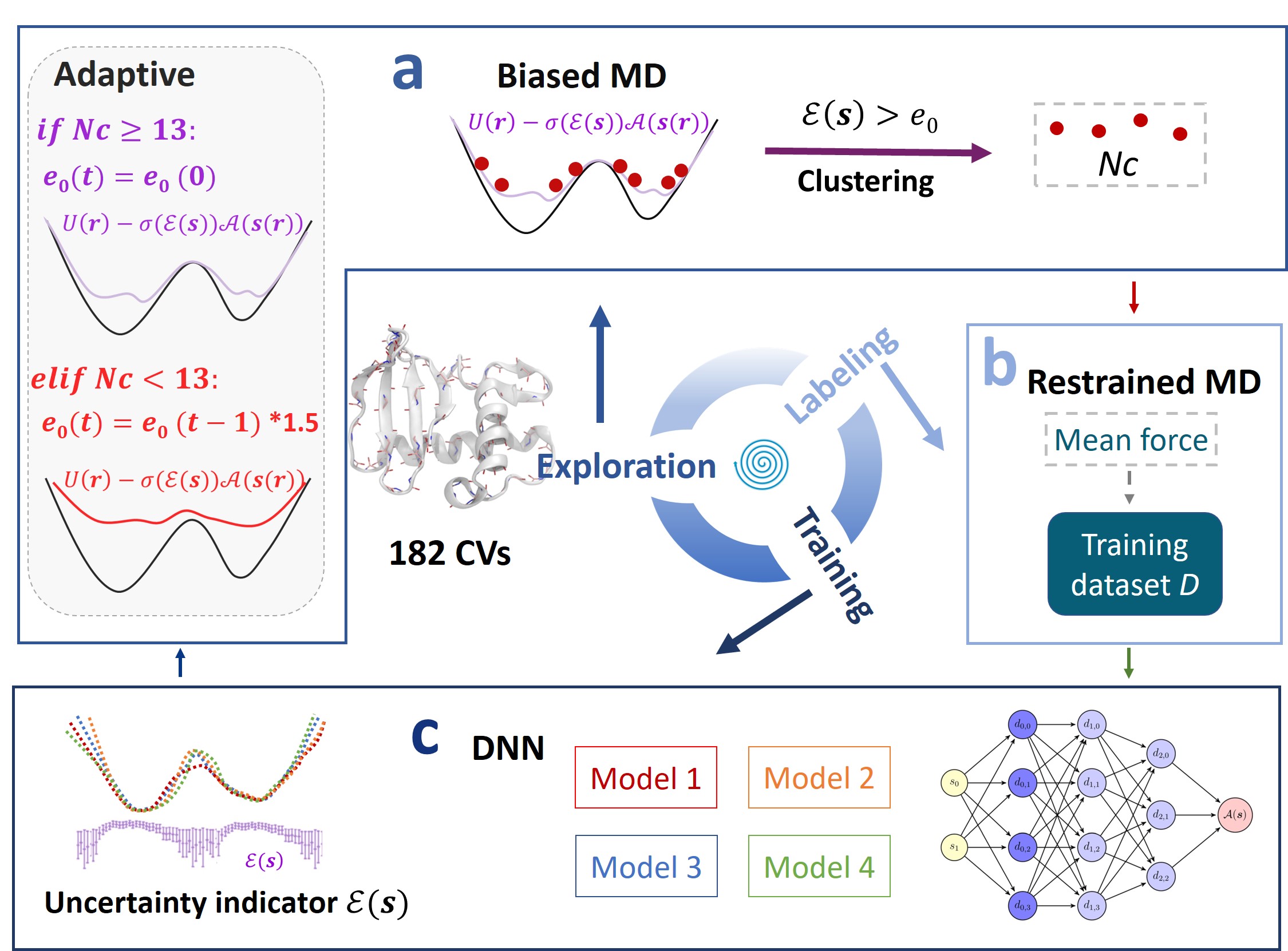}
\caption{{\bf The workflow of the adaptive RiD.}
Adaptive RiD iteratively and automatically promotes exploration, labeling, and training steps.
(a) Biased MD is used in the exploration step and the visited CV values with large uncertainty, viz.~$\me (\bm s) > e_0$, are proposed for labeling. 
The proposed CVs are then clustered into $Nc$ clusters, one set of CV values is randomly selected from each cluster for labeling.
An adaptive strategy is used at each iteration by changing the uncertainty levels according to the number of clusters $Nc$.
In detail, if $Nc$ is less than 13, then the level $e_0$ is multiplied by 1.5 and $e_1 = e_0 + 1$, otherwise the same levels as the initial values are used (gray panel).
(b) The mean forces evaluated by the restrained MD simulation are used as labels to train the DNN models.
(c) Four DNN models are trained by using different random initial parameters and the uncertainty indicator $\me(\bm s)$ is defined as the standard deviation of the force predictions from this ensemble of DNN models.
}
\label{fig:workflow}
\end{figure}

\section{Results}

The adaptive RiD method runs in iterations and each iteration consists of three steps (Fig.~\ref{fig:workflow}): {\bf exploration, labeling}, and \textbf{training}. 
In the exploration step, the sampling of the configuration space is enhanced by a biasing force defined in the CV space. 
The biasing force is switched on (or off) according to the value of an uncertainty indicator that is defined as the standard deviation of the force predictions from an ensemble of DNN models. 
The explored configurations are selected firstly according to the uncertainty indicator, then by a clustering argument, so that the a diversified subset of the sampled configurations with large model prediction error is selected. 
In the labeling step,  we calculate the mean force of every selected configuration as its label.
In the training step, we train an ensemble of DNN models with random and independent parameter initializations.
The section {\bf Methods~The~adaptive~RiD~method} provides more details on the adaptive RiD method.

Four representative systems are used to illustrate the performance of adaptive RiD.
First we use the peptoid system to compare adaptive RiD with its original version.
Then, we move to a more complex situation, the peptoid trimer with relatively large energy barriers, where we construct a 9-dimensional free energy landscape and compare adaptive RiD with other methods.
Next, we study a classical protein system, chignolin.
Using the 18 backbone dihedral angles as the CVs, we can employ adaptive RiD to efficiently sample the folding and unfolding states of chignolin.
Finally, we use adaptive RiD to help refine the protein structures of three targets in CASP13 using more than 100 CVs.

\begin{figure}
\centering
\includegraphics[width=.95\linewidth]{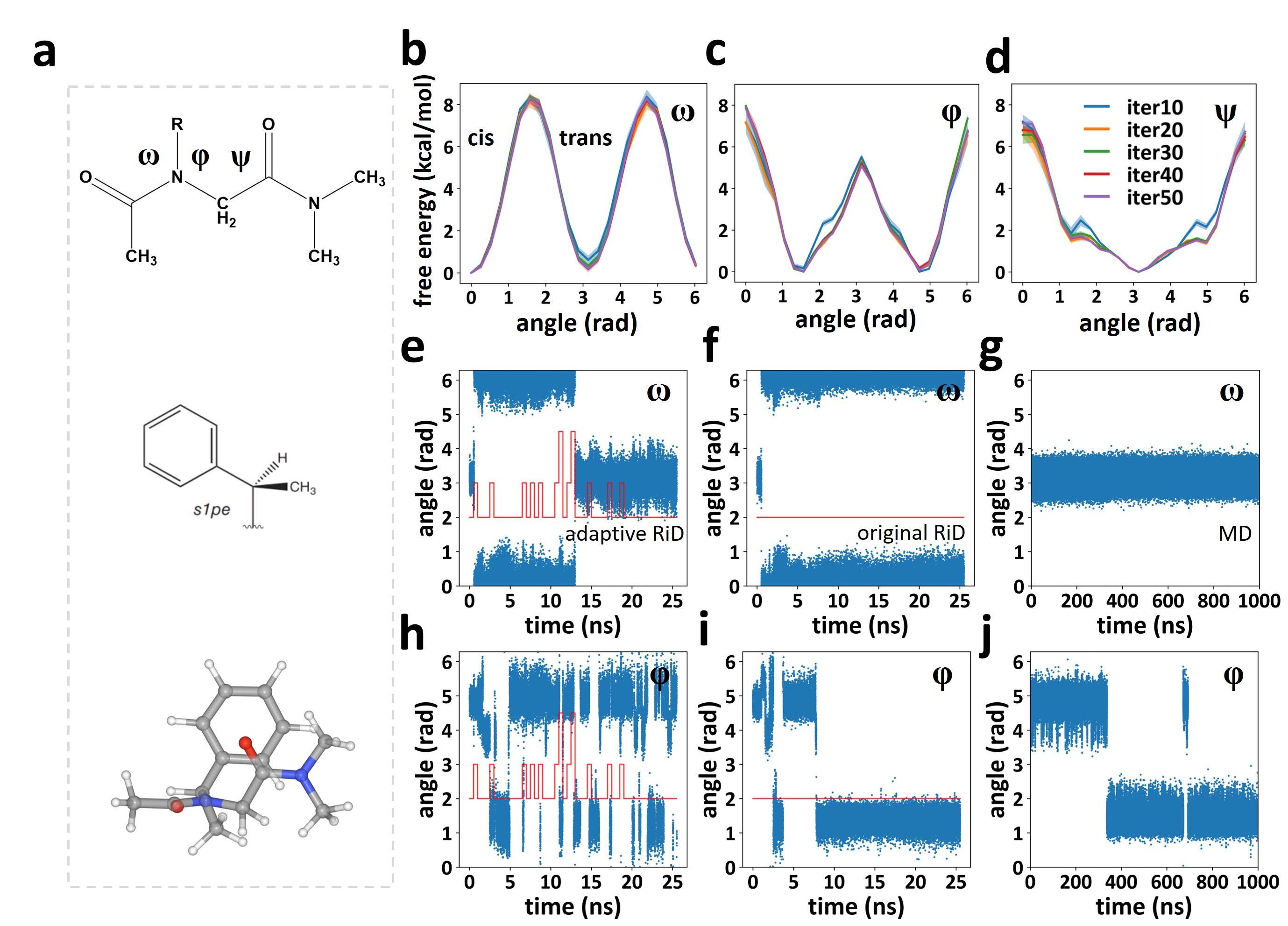}
\caption{{\bf The accuracy and efficiency of adaptive RiD.}
(a) Backbone plot, side chain plot, and 3D structure of the peptoid s1pe. 
The 3D structure is shown in Ball and Stick representation and colored by atomic types: nitrogen (blue), oxygen (red), hydrogen (white), and carbon (grey).
Free energy curve of each CV of the peptoid s1pe, $\omega$ (b), $\phi$ (c) and $\psi$(d), at different iterations (10th, 20th, 30th, 40th, 50th). 
The transitions of angles for the first walker of an adaptive RiD simulation (e, h), the first walker of an original RiD simulation (f, i), and a brute-force MD simulation (g, j) as a function of the simulation time.
Blue dots represent the torsion angles of $\omega$ (e-g and $\phi$ (h-j) of s1pe.  
The values of uncertainty level $e_0$ in adaptive RiD and original RiD simulations are drawn in red lines.}
\label{fig:peptoid-trans}
\end{figure}

{\bf Peptoid.}    Peptoids are a family of synthetic oligomers composed of protein-like, poly-glycine backbones with side chains (R) attached to the amide nitrogen atoms rather than their $\alpha$-carbons~\cite{ducheyne2015comprehensive,sun2013peptoid} (Fig.~\ref{fig:peptoid-trans}a). 
Recently, peptoids receive growing interest due to their resemblance to peptides, with applications such as bioactive peptide mimics~\cite{Mojsoska2015Structure} and targeted binders~\cite{li2010blockade,luo2013abeta42}. 
Despite a great amount of efforts~\cite{mirijanian2014development,mukherjee2015insights,weiser2017molecular,weiser2019cgenff}, efficient sampling of peptoid configurations remains challenging and is far from being routine due to the high dimensionality and the large energy barriers in the space of torsion angles.

In our work, we firstly consider a simple peptoid, s-(1)-phenylethyl (s1pe), which is obtained from Weiser's work~\cite{weiser2019cgenff}. The adaptive RiD simulation is conducted for 51 iterations, which corresponds to a 25.5-ns biased MD simulation for each walker (See {\bf Methods~Peptoid} for details).
The calculated FES is fully converged within 40 iteration, as shown in Fig.~2b-d.
The error of FES estimated by the maximal standard deviation of four DNN models is 0.31 kcal/mol.
Remarkably, the free energy difference between the trans ($\omega\sim$180${}^{\circ}$) and cis ($\omega\sim$0${}^{\circ}$) conformations is 0.18 $\pm$ 0.07~kcal/mol, which agrees quite well with experiment (0.14 kcal/mol)~\cite{gorske2009new}.
We show that the FES is consistent with previous simulation results~\cite{weiser2019cgenff} (Supplementary~Figure~1) and with that calculated by the REMD method (Supplementary~Figure~2).

We demonstrate the improvement of adaptive RiD upon the original RiD and brute-force MD in terms of sampling efficiency by studying the transition events.
As shown in Fig.~\ref{fig:peptoid-trans}g, j, a 1-$\mu$s-long brute-force MD simulation exhibits very few transitions of the torsion angles $\omega$ and $\phi$.
Transitions happen in a 25.5-ns-long original RiD simulation (Fig.~\ref{fig:peptoid-trans}f, i), but less frequently than those in an adaptive RiD simulation of the same length (see also Supplementary~Table~1).
The number of explored transition states ($\omega \in [3\pi/8, 5\pi/8] \cup [11\pi/8, 13\pi/8]$) is 725 and 266 for adaptive and original RiD, respectively.
\redc{While, no transition states can be seen from the REMD simulations (Supplementary~Figure~2).}
In conclusion, adaptively changing the uncertainty level $e_0$ (red lines in Fig.~\ref{fig:peptoid-trans}e-j, see the definition in \textbf{Methods~The~adaptive~RiD~method}) helps accelerate the exploration of conformation space.

\begin{figure}
\centering
\includegraphics[width=.9\linewidth]{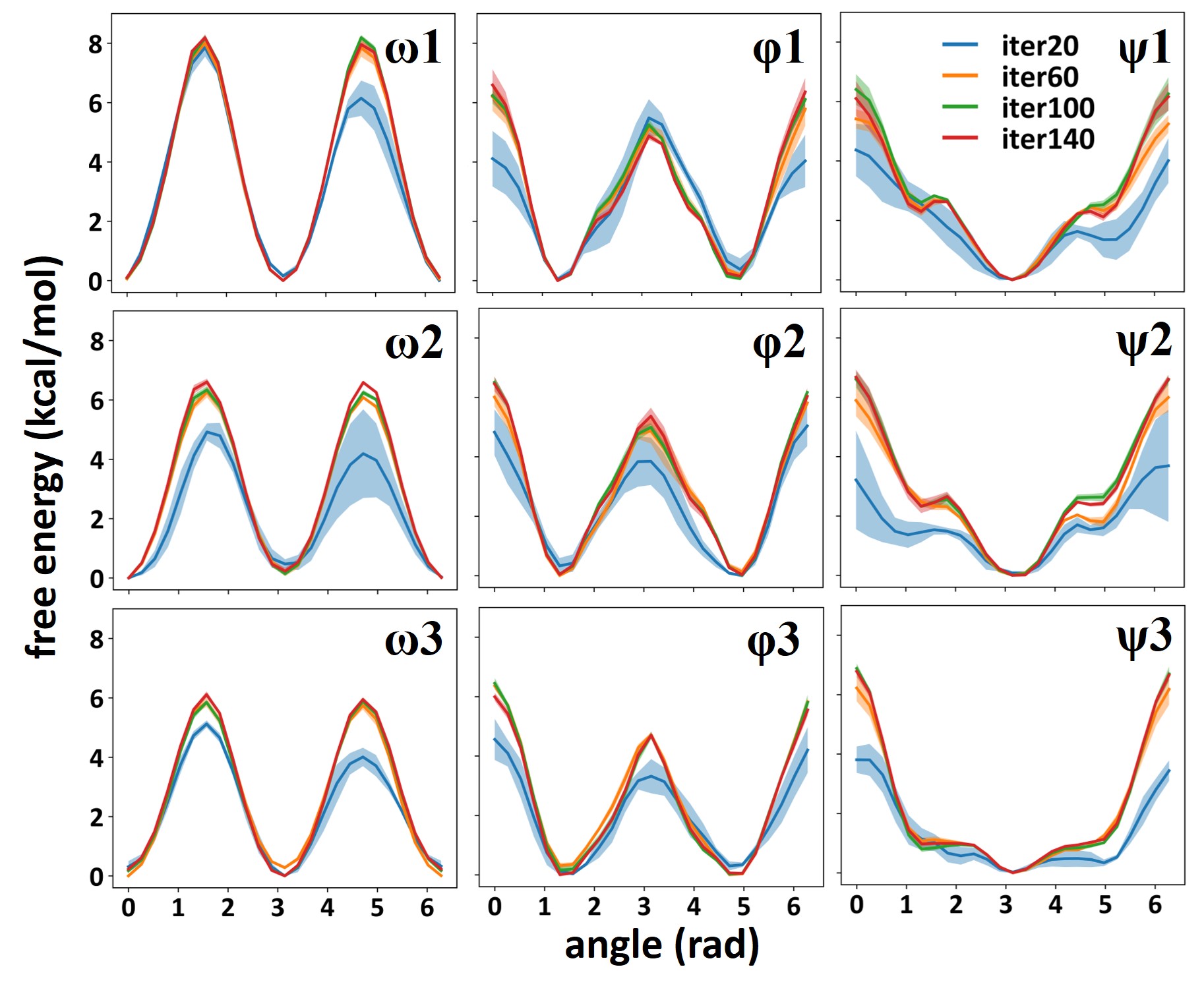}
\caption{{\bf Free energy curves of the peptoid trimer (s1pe)${}_3$.} 
Free energy curve of each CV of the peptoid trimer (s1pe)${}_3$.
The free energy curves derived from different iterations (20th, 60th, 100th, 140th) are drawn to assess the convergence of adaptive RiD.
Error bars are calculated based on four DNN models. 
}
\label{fig:peptoid}
\end{figure}

Furthermore, we consider a more complex system, the peptoid trimer (s1pe)${}_3$, which exhibits high free energy barriers in the space of nine torsion angles $\omega_1$, $\phi_1$, $\psi_1$, $\omega_2$, $\phi_2$, $\psi_2$, $\omega_3$, $\phi_3$, and $\psi_3$.
A previous study showed that it is hard to observe transitions of torsion angle $\omega$ and $\phi$ in simulations with explicit solvent~\cite{weiser2019cgenff}.
In our simulation, 12 walkers are used and each lasts 140 iterations.
The biased MD simulation in each iteration lasts 2 ns (see {\bf Methods~Peptoid} for details). 
As can be seen from Supplementary~Table~2, we observed tens to hundreds of transitions of the six torsion angles ($\omega_1$, $\omega_2$, $\omega_3$, $\phi_1$, $\phi_2$, $\phi_3$), much more than what was observed in a previous work~\cite{weiser2019cgenff} which performed a 85-ns brute-force MD simulation at 950K.
The overall trend of the transition frequency at different angles is qualitatively consistent with the result of the previous MD simulations at high temperature~\cite{weiser2019cgenff}.

We use a Markov Chain Monte Carlo (MCMC) scheme to project the FES on fewer CVs (see {\bf Methods~MCMC~simulations} for details).
Fig.~\ref{fig:peptoid} exhibits an energy barrier of more than 8 kcal/mol in $\omega_1$. 
The free energy differences between metastable states can be estimated fairly accurately at the early stage of the adaptive RiD simulation (20th iteration). 
In fact the FES has almost converged after 60 iterations (about 1440 ns).
Note that the FES in the space of $\psi$3 is quite flat over a large region.
This indicates that the carboxy termini ($\phi$3, $\psi$3) is quite flexible, which confirms the conclusion in a previous study~\cite{weiser2019cgenff} obtained by high-temperature simulations.

As a comparison, we calculated the free energy surface of peptoid trimer by replica exchange molecular dynamics (REMD) simulations (see \textbf{Methods~Peptoid~trimer} for more details and \textbf{Methods~REMD~Simulations~of~Peptoid} for more discussion). 
The FES calculated by REMD is sensitive to the setting of temperature range. 
Even with the optimal temperature range, the transition states of the $\omega$ and $\phi$ angles are not properly sampled by REMD, so the free energy profile is hard to converge at the transition states. 
This difficulty can be avoided by using adaptive RiD.
BEMetaD and PBMetaD simulations are also conducted as another comparison.
The same set of CVs and 9 replicas are used under different conditions  (detailed in {\bf Methods~Peptoid~trimer}).
The free energy curve of each angle essentially converges after 160 ns for each replica (see Supplementary~Figure~3), accumulating to a total of 1440~ns of simulation time, with the same time as adaptive RiD.
The free energy curves of BEMetaD, PBMetaD, adaptive RiD and REMD are in good agreement (see Supplementary~Figure~4),
and their errors with respect to the best converged REMD (300-680K, 400~ns each replica) are similar (see Supplementary~Table~3).
Overall, adaptive RiD has comparable efficiency and accuracy with BEMetaD and PBMetaD.


\begin{figure}
\centering
\includegraphics[width=.95\linewidth]{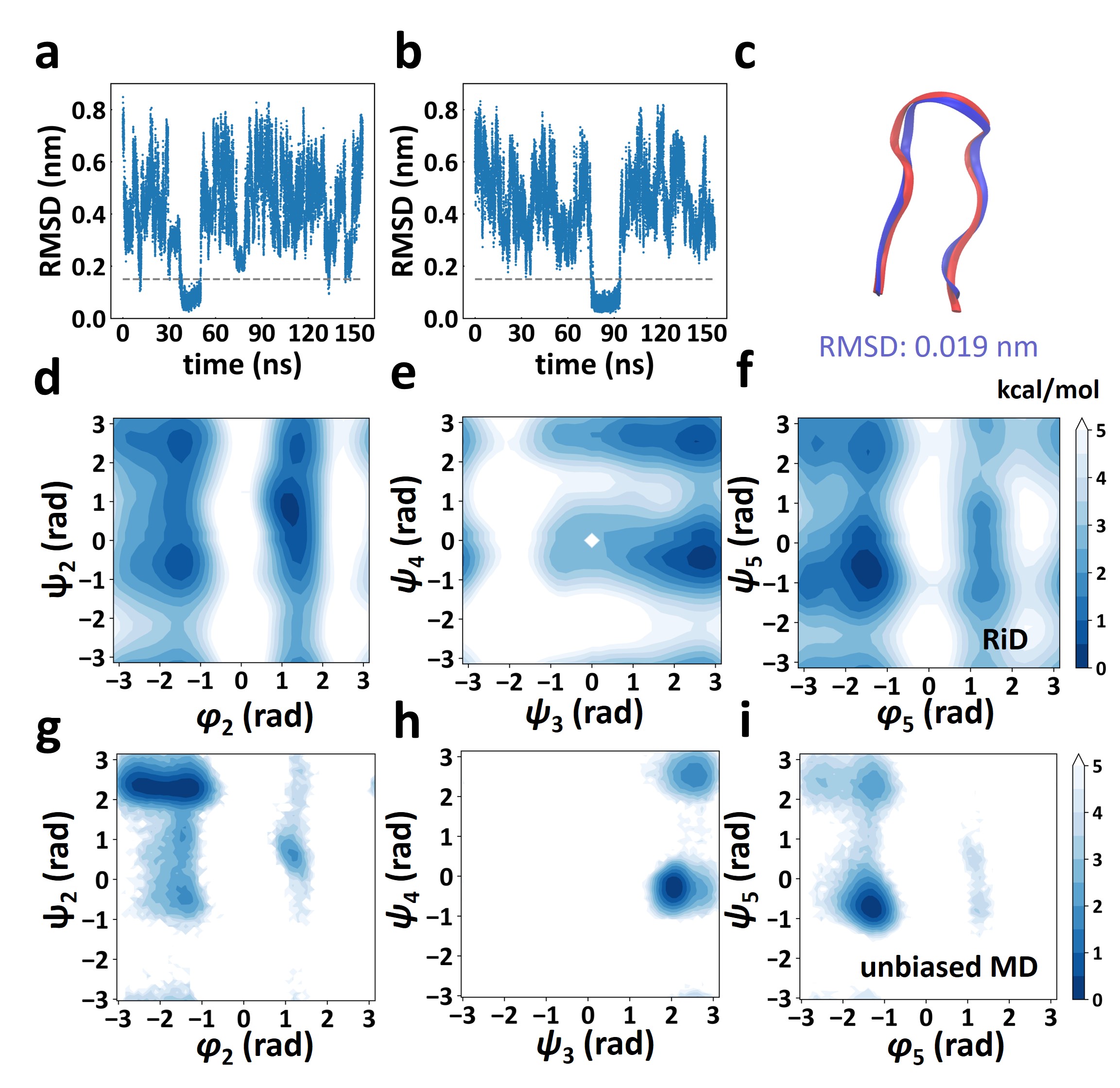}
\caption{{\bf Folding and unfolding of the protein chignolin.} (a, b) Time series of the C$\alpha$ RMSD from the reference structure for chignolin for two walkers in the adaptive RiD. 
Structures with RMSD less than 0.15 nm (grey dash line) are defined as folded states. 
(c) The folded structure obtained from adaptive RiD (blue) superimposed on the reference structure (red).
The C$\alpha$ RMSD is 0.019 nm. 
(d-f) Free energy surfaces (kcal/mol) of $\phi_2$ and $\psi_2$ (d), $\psi_3$ and $\psi_4$ (e), $\phi_5$ and $\psi_5$ (f) from adaptive RiD. 
(g-i) The corresponding free energy surfaces obtained from a 100-$\mu$s unbiased MD trajectory from the work of  Lindorff-Larsen and their coworkers~\cite{lindorff2011fast}.}
\label{fig:chignolin}
\end{figure}

{\bf Chignolin.}    
Folding of small proteins serves as a good benchmark for first-principles-based methodologies.
Here, we choose the artificial protein chignolin, which shows a $\beta$-hairpin structure (with 10 residues), as an example. 
There are both extensive experimental ~\cite{honda2008crystal}, 
and simulation studies~\cite{lindorff2011fast,kuhrova2012force,zhang2015folding,miao2015accelerated,shaffer2016enhanced} for this system.
In particular, using the Anton machine, an MD simulation lasting longer than 100 $\mu$s has been conducted~\cite{lindorff2011fast}, and a folding and unfolding time of about 0.6 $\mu$s and 2.2 $\mu$s, respectively, were observed. 

The adaptive RiD uses 18 backbone dihedral angles as the CVs with 12 walkers in parallel.
The initial configurations are randomly extended conformations and 31 iterations are conducted.
In each iteration, a 5-ns biased MD simulation is performed for each walker (see {\bf Methods~Chignolin} for details). 
As shown in Fig.~\ref{fig:chignolin}a,b, the folding and unfolding events are observed within a few iterations. 
The minimum C$\alpha$ RMSD is 0.019 nm (Fig.~\ref{fig:chignolin}b,c). 
Both the folding and unfolding rates are observed to be equal to 4.30~$\mu s^{-1}$.
\redc{Therefore, adaptive RiD efficiently samples the folding and unfolding events of chignolin without prior knowledge on the native state, in comparison to the VES scheme~\cite{shaffer2016enhanced}, which needs a target distribution to lead the sampling.} 
BEMetaD and PBMetaD simulations are also conducted to compare with adaptive RiD.
The same set of CVs is used and different working parameters are considered  (see {\bf Methods~Chignolin} for details).
The folding and unfolding rates are reported in Supplementary~Table~5.
Overall, the adaptive RiD and all variants of MetaD successfully fold chignolin, but only the adaptive RiD and PBMetaD can escape from the native state. 
This implies that adaptive RiD and PBMetaD are more suitable for larger proteins that have deep metastable states on the free energy landscape.

In Fig.~4 we compare the FES calculated by adaptive RiD (Fig.~\ref{fig:chignolin}d-f) with a 100-$\mu$s unbiased MD trajectory (Fig.~\ref{fig:chignolin}g-i). 
We find qualitative agreement between the two results, and adaptive RiD samples a broader region in the conformation space using a much shorter exploration time ($1.8\mu$s).
Moreover, the result also agrees well with the ones obtained by the VES scheme in a previous study~\cite{shaffer2016enhanced}.

{\bf Protein structure refinement.}    
In the past few years, a significant progress has been made on the protein folding problem.
Of particular importance is the reported remarkable achievement of AlphaFold2~\cite{alphafold2}.
However, for 47\% of all targets or 37\% of those with less than 100 residues, the structures predicted by AlphaFold2 in CASP14 have Global Distance Test-High Accuracy (GDT-HA{)}~\cite{zemla2003lga} scores less than 75.
GDT-HA scores range from 0 to 100, and a higher score means a higher similarity between the predicted and native structures.
For these structures, an additional structure refinement procedure might generate better predictions.
Several refinement protocols have been reported in the literature~\cite{raval2012refinement,feig2016protein,heo2018experimental,park2018protein,park2019high,heo2019driven}, including those based on MD simulations~\cite{feig2016protein,heo2018experimental}.
In the refinement category of the CASP13 competition, in which the goal is to generate a better predicted structure using the given decoys as starting points, the FEIGLAB, using MD simulations, achieved the best results with an average improvement of GDT-HA scores of about 3.99 units.
\redc{However, as the authors noted, sufficient sampling remains challenging for some proteins because of the kinetic barriers~\cite{heo2019driven}. } 

\begin{figure}
\centering
\includegraphics[width=1\linewidth]{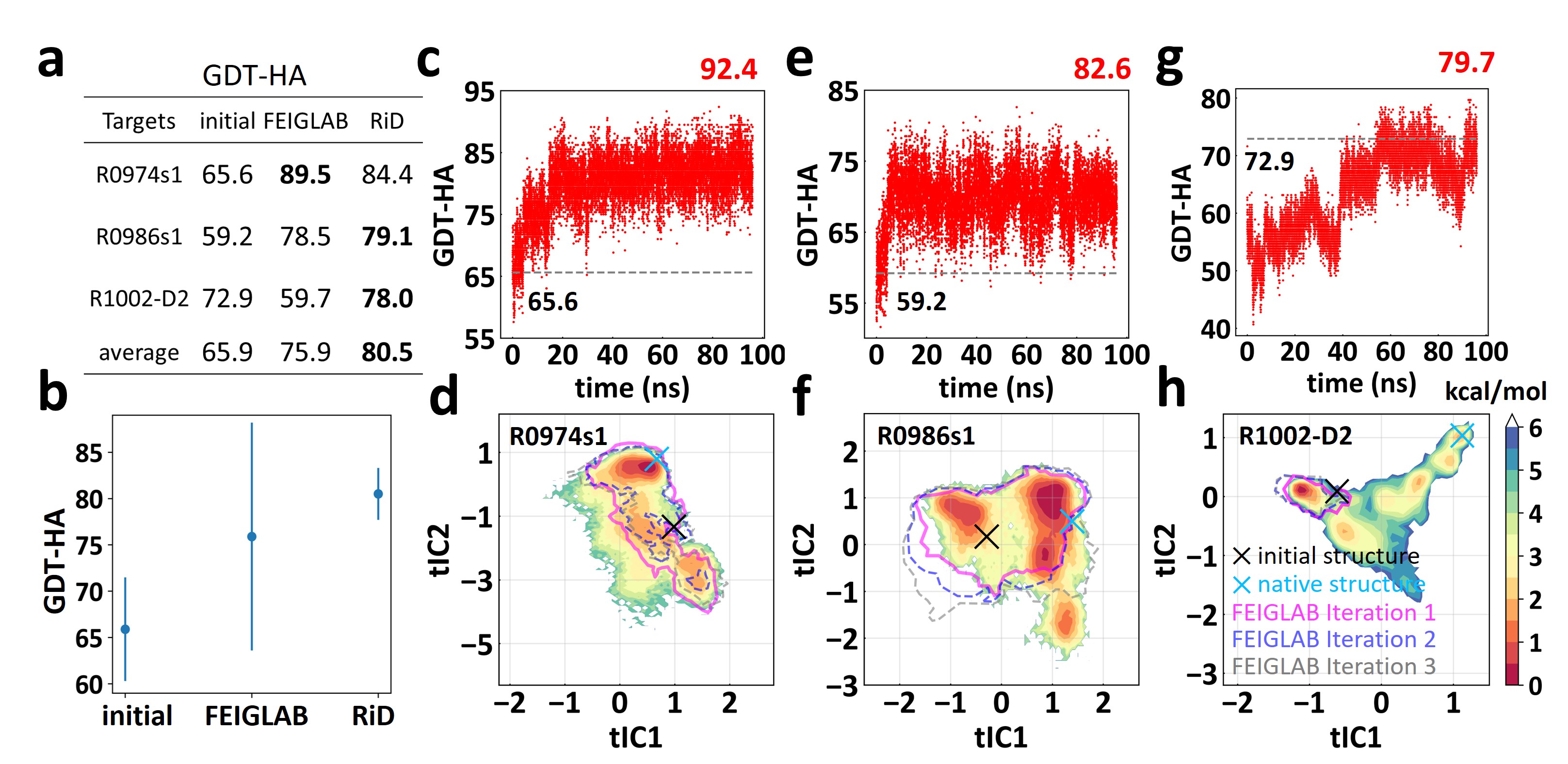}
\caption{{\bf Protein structure refinement of three targets, R0974s1, R0986s1 and R1002-D2.} The values (a) and the plot (b) of GDT-HA scores of the initial structures, the refined structures of FEIGLAB and the refined structures of adaptive RiD. The error-bar in panel b is the standard deviation of these three targets. (c, e, g) Time series of the GDT-HA score of one representative trajectory for each target, R0974s1 (c), R0986s1 (e), and R1002-D2 (g).
(d, f, h) Density plots of the two tIC coordinates, used in the MSM analysis in Heo's work~\cite{heo2019driven}, along the adaptive RiD trajectories of R0974s1 (d), R0986s1 (f), and R1002-D2 (h).
The explored regions in Heo's work~\cite{heo2019driven} for these targets are drawn as magenta line (iteration 1), blue dash line (iteration 2) and grey dash line (iteration 3) respectively. }
\label{fig:refine1}
\end{figure}

Here, we choose three CASP13 targets, R0974s1, R0986s1, and R1002-D2, which were analyzed in detail in the work of Heo et al~\cite{heo2019driven}.
We use 8 walkers and perform 16 iterations. 
Each iteration lasts 6 ns, so a total of 96-ns biased MD simulation is conducted for each walker. 
136, 182, and 116 dihedral angles of these three targets are used as CVs respectively. In addition, to prevent transitions to the unfolded states, we apply flat-bottom harmonic restraint potential to each C$\alpha$ atom~\cite{heo2019driven} (see {\bf Methods~Protein~structure~refinement} for details).
The refined structure is constructed by averaging 200 structures with the lowest RWplus values~\cite{zhang2010novel} along the adaptive RiD trajectories.
The effectiveness of the adaptive RiD in terms of the improvement in GDT-HA score is reported and compared with the results of FEIGLAB in Fig.~\ref{fig:refine1}a, b.
The three targets have initial GDT-HA scores of  65.6, 59.2 and 72.9, respectively, and are all refined to beyond 78.0 under adaptive RiD scheme.
Special attention is drawn to R1002-D2, the score of which is improved to 78.0 after the refinement, but  is significantly worse than the score by FEIGLAB~\cite{heo2019driven}.
The average improvement in GDT-HA score is 14.6 units.  The standard deviation is 2.8. This low value shows the robustness of adaptive RiD.

To see how the protein structures are explored by the adaptive RiD, we calculated the GDT-HA scores along the adaptive RiD trajectories and found the best sampled structures with GDT-HA scores of 92.4, 82.6, and 79.7, respectively (Figs.~\ref{fig:refine1}c, e, g).
Notice that the structures with high scores are quickly reached within 20 ns in the cases of R0974s1 and R0986s1. 
In contrast, for R1002-D2, the GDT-HA score quickly drops in the beginning of the simulation, and then transition to a better score after more than 40 ns.
To further compare the conformational space explored by adaptive RiD with that sampled by FEIGLAB, we project the trajectories onto the two time-structure independent component (tIC) coordinates used in the Markov state model (MSM) analysis of the FEIGLAB study (Fig.~\ref{fig:refine1}d, f, h).
We find that, compared with the 2-$\mu$s-long brute-force simulations in the FEIGLAB sampling protocols, comparable conformation spaces are sampled by the 768-ns-long adaptive RiD simulations of R0974s1 and R0986s1, and a much larger conformation space is sampled in the case of R1002-D2.
Indeed, for R1002-D2, the FEIGLAB simulation only found one metastable state out of the three states that were investigated in their MSM study, but adaptive RiD finds all  three, including the native state.

\begin{figure}
\centering
\includegraphics[width=1.0\linewidth]{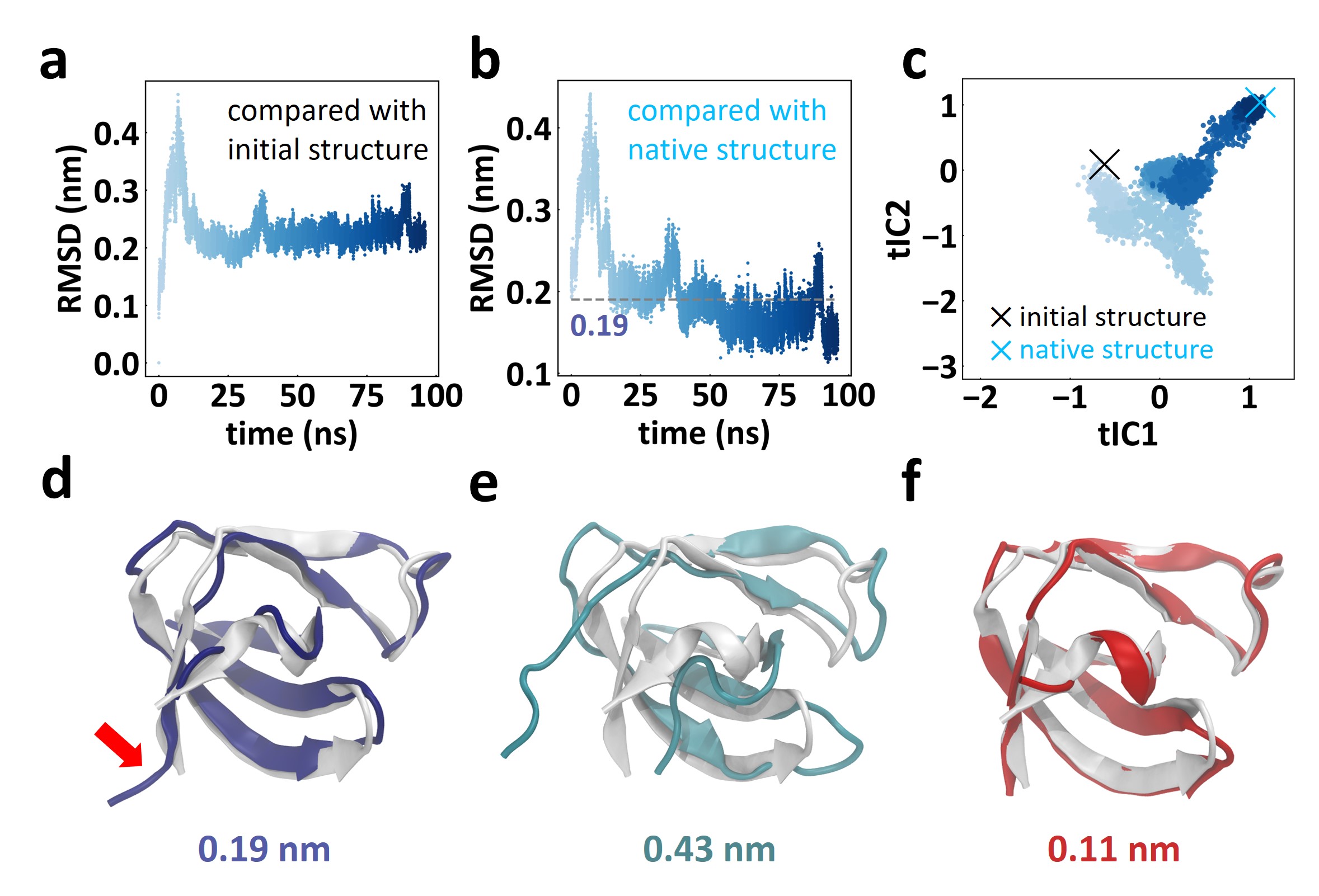}
\caption{{\bf The detailed analysis of  the target R1002-D2.}
(a-b) Time series of the RMSD between the structures along a representative trajectory and the initial structure (a), or native structure (b).
(c) One representative adaptive RiD trajectory projected on the two tIC coordinates.
Each circle represents one snapshot and is colored at different times from light blue to dark blue in panels a-c.
(d-f) The native structure (white) superposed on the initial structure (d, blue), the unfolded structure (e, cyan), and refined structure (f, red), respectively. 
RMSDs of the initial structure, unfolded structure and refined structure compared with the native structure are shown in panel d-f, respectively.
}
\label{fig:refine2}
\end{figure}

To see how adaptive RiD finds the native state of R1002-D2 from a conformational change  perspective, we perform a detailed analysis (Fig.~\ref{fig:refine2}).
According to Figs.~\ref{fig:refine2}a-c, it is clear that the initial structure has a high similarity to the native structure, but during the adaptive RiD simulation, the walker first goes to a meta-stable state that is distant from both the initial structure and the native structure, and it then falls into the basin of the native structure.
In detail, as shown by the red arrow in Fig.~\ref{fig:refine2}d, there is a register-shift error of about 0.57 nm in the N-terminal $\beta$-strand of the initial stricture. 
Then, during the simulation, the hydrogen bonds between the $\beta$-strands break at first and lead to a large conformational change with more than 0.4 nm in RMSD compared with both the initial model and the native structure (Fig.~\ref{fig:refine2}a, b, e). 
It is only after this large conformational change that the correct $\beta$-strand are seen and a more accurate structure with RMSD of about 0.11 nm is found (Fig.~\ref{fig:refine2}f). 
Therefore, we conclude that the success of adaptive RiD is attributed to the ability of crossing over the high free energy barrier along the transition path to the native state. 
In contrast, the same barrier forbids the FEIGLAB protocol from discovering the transition path.

\section{Discussion} 
With the ability to handle a large number of CVs, RiD is very powerful in exploring the configuration space of atomistic systems.
The added feature of adaptively changing the bias helps to facilitate the escape from metastable states and crossing large energy barriers.
Moreover, adaptive RiD supports a multi-walker scheme, and it is very much suited for parallelization.
It also opens up new possibilities, such as free energy calculation of the binding of peptides or intrinsically disordered proteins to proteins, the study of the ensemble nature of allostery, and structure optimization in materials science.

Several improvements may further enhance the performance of adaptive RiD.
For instance, for {\it ab initio} folding of large proteins, the space of dihedral angles might be too large, and a better set of CVs might be needed.
For the protein refinement task, with the help of quality assessment~\cite{jing2020improved} of protein tertiary structures, we 
may choose to include only the dihedral angles  in the residues with poor quality as the CVs to reduce computational cost.
We leave these possibilities to the future.

\section{References}

\section{Acknowledgements}
The work of D. W., L. Z., and W. E is supported in part by a gift from iFlytek to Princeton University. 
The work of H.W. is supported by the National Science Foundation of China under Grant No.11871110 and Beijing Academy of Artificial Intelligence(BAAI).
The work of L.Z.  is also supported by the DOE Center of Chemistry in Solutions and at Interfaces (CSI) through  Award DE-SC0019394.

\section{Author Contributions Statement}
D.W, L.Z., H.W. and W.E conceptualized the research; 
D.W., Y.W. and J.C. conducted the research and performed data analysis; D.W. L.Z., H.W. and W.E drafted the paper. All authors commented and revised the paper.

\section{Competing Interests Statement}
The authors declare no competing interests.
\\[30pt]

\section{Methods}
\textbf{The adaptive RiD method.} We consider $N$ atoms in a canonical ensemble,  whose potential energy is denoted by $U(\bm r)$, $\bm r = (\bm r_1, \dots, \bm r_N)$.
Given $M$ predefined CVs, denoted by $\bm s$, the free energy defined on the CV space is denoted by $\ma(\bm s)$.
A deep neural network (DNN), denoted by $\ma (\bm s, \bm w)$, is used to model the free energy, with $\bm w$ being the network parameters.
For a given set of CV values, the mean forces $\bm F (\bm s) = -\nabla_{\bm s} \ma(\bm s)$ are evaluated by restrained MD simulations~\cite{maragliano2008single}, through adding a harmonic potential between the target and the instantaneous CVs, and estimate $\bm F$ by the averaged restraining force.
The mean forces $\bm F(\bm s)$ are then used as labels to train the DNN models (Fig.~\ref{fig:workflow}b, c). 


To guarantee the accuracy of the neural network approximation of $\ma$, one should have an adequate training dataset $D$. 
How to efficiently explore the CV space and how to select the explored data points for labeling constitute the key components of RiD.
The key quantity that helps for both purposes is the uncertainty indicator $\me(\bm s)$, which is defined as the standard deviation of the force predictions from a small ensemble of DNN models with the same network architecture but trained with different randomly initialized parameters.
This ensemble of models typically gives rise to consistent predictions of the mean force in regions well covered by $D$, and scattered predictions in regions poorly covered by $D$ (Fig.~\ref{fig:workflow}c).

As shown in Fig.~\ref{fig:workflow}a, one biases the potential energy $U(\bm r)$ by the current approximation of the FES, so that the system is encouraged to escape free energy minima and  explore a broader region in the CV space.
Since the approximation of FES is of high (or low) accuracy in regions that have been (or have not been) adequately explored, the bias is switched on (or off) according to the value of the uncertainty indicator $\me(\bm  s(\bm r))$:
\begin{linenomath*}
\begin{eqnarray}\label{eqn:ab}
  \tilde {\bm {f}}_i(\bm r) = -\nabla_{\bm r_i} U(\bm r) + \sigma(\me(\bm s(\bm r))) \,
  \langle \nabla_{\bm r_i}\ma (\bm s(\bm r), \bm w) \rangle,
\end{eqnarray}
\end{linenomath*}
where the biasing potential $- \langle \ma (\bm s(\bm r), \bm w)\rangle$ is the mean value of the predictions of the predefined ensemble of DNN models, and $\sigma(\cdot)$ is a fixed switching function that varies smoothly between two  uncertainty levels $e_0$ and $e_1$. 
When the uncertainty value is smaller than $e_0$, the accuracy of the FES approximation is adequate and the bias is switched on with $\sigma(\me) = 1$.
When the uncertainty value is  larger than $e_1$, the bias is switched off ($\sigma(\me) = 0$) and the dynamics of the system falls back to the one governed by the original potential energy $U(\bm r)$.
The switching function is defined by
\begin{align}
\label{eqn:epsilon}
 \sigma(\epsilon)=
    \begin{cases}
             1, & \epsilon<\epsilon_0 \\
             \frac{1}{2}+\frac{1}{2}\cos{(\pi \frac{\epsilon-\epsilon_0}{\epsilon_1-\epsilon_0})}, & \epsilon_0 <\epsilon < \epsilon_1 \\
             0, &\epsilon > \epsilon_1
    \end{cases}
\end{align}
During the exploration, the visited CV values are tested by the uncertainty indicator and proposed for labeling if the uncertainty is relatively large, viz.~$\me (\bm s) > e_0$ (Fig.~\ref{fig:workflow}a).
The proposed CVs are first clustered by the agglomerative clustering algorithm, and one CV values is randomly selected from each cluster for labeling.
This reduces the number of samples in the CV space needed for labeling without affecting the representativeness of these samples.
This CV selection procedure is applied after the exploration step in every adaptive RiD iteration.
Finally, theselected CV values and the associated labels are added to the training dataset $D$, from which a new ensemble of DNN models are trained. 
The new iteration is started over  with the updated model ensemble.

The uncertainty levels $e_0$ and $e_1$ are two crucial parameters in the RiD method.
A large level would encourage exploration at the cost of lowing the accuracy of the FES approximation.
A small level gives more accurate approximation of the local FES, but the system is easily trapped by deep energy wells.
Here we adaptively change the uncertainty levels at each iteration.
In detail, the explored configurations are clustered by the agglomerative clustering algorithm. 
If the number of clusters is less than some fixed value $K_0$ ($K_0=13$ for this work), then the level $e_0$ is multiplied by 1.5 and $e_1 = e_0 + 1$, otherwise the same levels as the initial values are used. If $e_0$ is more than 8 times larger than its initial value, then both levels are reset to their initial values (Fig.~\ref{fig:workflow}a, gray panel). We refer to this as the adaptive RiD.
The comparison of the efficiency of the original RiD and adaptive RiD us given in the following parts.

\textbf{Simulation protocol.} We use the GROMACS/2019.2 software package~\cite{abraham2015gromacs} with a modified version of the PLUMED/2.5.2 plugin~\cite{tribello2014plumed} to conduct all MD simulations. 
The PLUMED package is modified to compute the DNN biasing force, viz., Eq.1.
We first minimize the energy of proteins using the steepest descent algorithm. Then the added solvent is equilibrated with position restraints on the heavy atoms of the protein.
Temperature is maintained using the V-rescale method~\cite{bussi2007canonical} with a relaxation time of 0.2~ps. 
We use the Parrinello-Rahman barostat~\cite{parrinello1981polymorphic} to keep the pressure at 1~bar with a time constant of 1.5~ps. 
The cutoff of electrostatic interactions and van der Waals interactions are both set to be 1~nm, and the particle mesh Ewald method~\cite{darden1993particle} is used to treat electrostatic interactions. 
All bonds involving H atoms are constrained by the LINCS algorithms~\cite{hess1997lincs}.

{\bf Peptoid.} In order to assess the accuracy and efficiency of adaptive RiD, we first study one example system, the peptoid with side chain s-(1)-phenylethyl (s1pe) which is obtained from Weiser's work~\cite{weiser2019cgenff}.
The molecules are solvated in (2.9 nm)${}^3$ dodecahedron boxes with 546 water molecules.
The final systems contain 1676 atoms.
All simulations are carried out with the CHARMM general force field (CGenFF) parameters developed for peptoids~\cite{weiser2019cgenff} and the TIP3P water model~\cite{jorgensen1983comparison} for the explicit solvent.
The temperature is maintained at 300 K.
After a conventional 1000 ns MD simulation is conducted,
12 conformations are randomly selected from the trajectory for the initial configurations of adaptive RiD.
Three torsion angles $\omega$ (C${}_\alpha$, C, N, C${}_\alpha$), $\phi$ (C, N, C${}_\alpha$, C) and $\psi$ (N, C${}_\alpha$, C, N) are chosen as CVs, i.e., $\bm s$ = ($\omega$, $\phi$, $\psi$).
The preprocessing operator is taken as $P (\omega, \phi, \psi) = (\cos(\omega), \sin(\omega), \cos(\phi), \sin(\phi), \cos(\psi), \sin(\psi))$, so the periodic condition of the FES is guaranteed.
12 walkers with different conformations of adaptive RiD are conducted using 500 ps biased MD simulations. 
The CV values along the MD trajectories are computed and recorded in every 0.5 ps. 
We assume no prior information regarding the FES, so brute-force simulations are performed for the 0th iteration step (we count the iterations from 0).
In each iteration, the recorded CV values in the region with high uncertainty are clustered using the agglomerative clustering algorithm with a distance threshold.
The distance threshold is chosen such that, for a conventional MD simulation, about 15 clusters are formed.
One CV value from each cluster is added to the training dataset $D$. 
Restrained MD simulations with spring constant 500 (kJ/mol)/rad${}^2$ are then performed to estimate the mean force.
Each restrained MD simulation is 100 ps long for both systems.
The CV values are recorded in every 0.1 ps along the restrained MD trajectories to estimate the mean forces.
The uncertainty levels of the reinforced dynamics are set to $e_0$ = 2.0 (kJ/mol)/rad, $e_1$ = 3.0 (kJ/mol)/rad and $e_{accept}$ = 2.0 (kJ/mol)/rad.
For RiD, the uncertainty levels $e_0$ and $e_1$ remain fixed, while for adaptive RiD, they are adaptively changed based on the number of clusters, $N_c$ = 13.
The DNN models contain four hidden layers of size (M1, M2, M3, M4) = (200, 200, 200, 200). 
Model training is carried out using the deep learning framework TensorFlow~\cite{abadi2016tensorflow} with the Adam stochastic gradient descent algorithm~\cite{kingma2014adam-a} with a batch size of $|B| = 128$.
The learning rate is 0.0006 in the beginning and decays exponentially according to $r_{l}(t)=r_{l}(0) \times d_{r}^{t / d_{s}}$, where $t$ is the training step, $d_r$ = 0.96 is the decay rate, and $d_s$ = $50 \times |D|/|B|$ is the decay step. 
The total number of training steps is $12,000 \times |D|/|B|$. 
Four DNN models with independent random initialization are trained in the same way to compute the uncertainty indicator. 
Details on the hyper-parameter tuning is explained in the final subsection of {\bf Methods A guideline for tuning the hyper-parameters of the adaptive RiD}. The whole procedure is conducted for 51 iterations.

\redc{For REMD simulations, a total of 20 different temperatures ranging from 300-430 K are generated using Temperature generator for REMD-simulations~\cite{patriksson2008temperature}.
The exchange time between two adjacent replicas is 2 ps and each replica lasts for 100 ns, giving 2 $\mu$s of simulation in total.
The average acceptance ratio is about 44\%.
The free energy curve of each angle is calculated from the trajectory at 300K and the convergence is assessed in Supplementary~Figure~2.}

{\bf REMD Simulations of Peptoid.}  
The REMD method runs an ensemble of MD simulations at a set of temperatures starting at the investigated temperature and ending at a high temperature that encourages the exploration in the conformation space. 
Intermediate temperatures are provided in the temperature range to make sure that the exchange probability between the replicas running at neighboring temperatures is significant. 
\emph{A priori} knowledge of the free energy surface is thus needed to properly set the highest temperature, which should be high enough to overcome the highest barrier of the free energy landscape, and be as low as possible to reduce the number of replica to save the computational cost. 
In the case of peptoid trimer, in which the \emph{a priori} knowledge is not available, a systematic of the free energy convergence with the temperature setting is needed. 

We report REMD simulations with temperature ranges of 300-430K, 300-560K and 300-680K here, which runs an ensemble of MD simulations at a set of temperatures starting at the investigated temperature and ending at a high temperature that encourages the exploration in the conformation space.
Wider temperature ranges are not shown because they need a smaller time-step to stabilize the MD simulation at higher temperature replicas and thus requires significantly more computational resources. 
The free energy curves derived from REMD with the temperature range 300-680K are the closest to adaptive RiD simulation among these conditions (Supplementary Figure 7, red lines and gray dash lines).
The shape of free energies obtained from REMD 300-430K simulations deviates from those from REMD 300-560K and 300-680K, especially the free energies against the angles $\phi_1$, $\phi_2$ and $\psi_2$ (Supplementary Figure 7, blue lines).
The time series of the peptoid trimer dihedral angles calculated on the continuous replicas that span different temperatures are shown in Supplementary Figure 8.
In addition, the total number of transitions of the six torsion angles $\omega_1$, $\omega_2$, $\omega_3$, $\phi_1$, $\phi_2$, $\phi_3$ of (s1pe)3 in different REMD simulations are shown in Supplementary Table 2.
We observe that there are tens of transitions in the REMD simulations of 300-430K and much more transitions are seen in the REMD simulations of 300-560K and 300-680K.
This indicates that higher temperatures make it easier for the algorithm to converge.
Therefore, 
Overall, a minimal temperature range of 300-560K is needed.
Even with temperature range 300-680K, the transition states of the $\omega$ and $\phi$ angles are not properly sampled by REMD, so the free energy profile is hard to converge at the transition states. 
There are two difficulties  in the REMD simulations: (1) a systematic and expensive investigation about the temperature ladder is needed to ensure the convergence of the free energy calculation, and (2) the free energy at the transition states of high energy barriers are not available.
These difficulties can be avoided by using adaptive RiD.

{\bf Peptoid trimer.} The molecules are solvated in (3.5 nm)${}^3$ dodecahedron boxes with 973 water molecules.
The final systems contain 3003 atoms.
Here, the adaptive RiD simulation is conducted for 12 walkers and 140 iterations.
All other parameters are the same as in {\bf Peptoid} except that the biased MD simulation lasts for 2ns.

\redc{For REMD simulations, a total of 27, 42 and 56 different temperatures ranging from 300-430 K, 300-560 K and 300-680 K, respectively, are generated using the Temperature generator for REMD-simulations~\cite{patriksson2008temperature}.
The exchange time between two adjacent replicas is 2 ps and each replica lasts for 100 ns, giving 2.7, 4.2 and 5.6 $\mu$s of simulations in total, respectively.
Three independent REMD simulations of 300-680K are conducted in order to obtain the error bar (Supplementary~Figure~6, red lines).
The average acceptance ratios are all about 40\%.
}
An additional REMD with each replica simulated for 400~ns at 300-680~K is conducted to quantitatively compare the accuracy of FES calculation by different methods (Supplementary~Figure~4).

For BEMetaD simulations, 9 dihedral angles are used as CVs.
Nine replicas are used and one for each collective variable.
Each replica lasts for 200 ns and the replicas are allowed to exchange every 20 ps.
The average acceptance ratio is about 20\%.
Three different sets of parameters are used.
For \textbf{BE0.2} which is derived from plumID:19.059, the parameters are [PACE=1000 HEIGHT=0.2 SIGMA=0.17 GRID\_MIN=-pi GRID\_MAX=pi BIASFACTOR=10.0 TEMP=300].
The parameters of \textbf{BE0.5} derived from plumID:21.014 are [PACE=500 HEIGHT=0.5 SIGMA=0.015 SIGMA\_MAX=0.6 SIGMA\_MIN=0.03 ADAPTIVE=GEOM GRID\_MIN=-pi GRID\_MAX=pi BIASFACTOR=10.0 TEMP=300].
The parameters of \textbf{BE0.8} are [PACE=500 HEIGHT=0.8 SIGMA=0.25 GRID\_MIN=-pi GRID\_MAX=pi BIASFACTOR=10.0 TEMP=300].
The free energy curve of each angle is calculated from the trajectories and  convergence is assessed in Supplementary~Figure~3.

For PBMetaD simulations, 9 dihedral angles are used as CVs.
9 replicas are simulated and each replica lasts for 200 ns.
The parameters are derived from Plumed Nest with ID plumID:21.014 [PACE=500 HEIGHT=0.5 SIGMA=0.015 SIGMA\_MAX=0.6 SIGMA\_MIN=0.03 ADAPTIVE=GEOM GRID\_MIN=-pi GRID\_MAX=pi BIASFACTOR=10.0 TEMP=300].
This PBMetaD setting is denoted by \textbf{PB0.5}.

{\bf Chignolin.} For chignolin, the crystal structure was first obtained from Protein Data Bank (PDB ID: 5AWL)~\cite{honda2008crystal}.
In order to achieve the initial unfolded conformations, one MD simulation in vacuum at 1000 K is conducted for 5 ns.
12 fully extended conformations are randomly chosen for the initial conformations of adaptive RiD.
They are solvated in a (4.2 nm)$^3$ dodecahedron box with 1622 water molecules and 2 sodium ions to neutralize the charge.
All simulations are carried out with the CHARMM22* force field~\cite{piana2011robust} and the TIP3P water model~\cite{jorgensen1983comparison}. The temperature is maintained at 340 K (in agreement with some previous work ~\cite{lindorff2011fast,shaffer2016enhanced}).
18 backbone dihedral angles are set as CVs.
12 walkers with different conformations of adaptive RiD are conducted for 31 iterations and the biased MD simulation lasted for 5 ns in each iteration.
The CV values are computed every 5 ps.
All other parameters are the same as in {\bf Peptoid}.
The reference structure of chignolin is generated from a short, conventional MD simulation that started from the X-ray structure (PDB ID: 5AWL).
This reference structure is a better reflection of the actual folded state of the protein in the CHARMM22* force field than the crystal structure.
Since the maximum RMSD between X-ray and the NMR (PDB ID: 2RVD) structures is about 0.18~nm, we define the folded state to be the structure with RMSD $<$ 0.19 nm from the reference structure, and the unfolded state to be the structure with RMSD $>$ 0.3 nm~\cite{paissoni2021determine}. 
The RMSD is smoothed along the trajectories with a sliding window of width 3~ns to eliminate the overestimate of folding/unfolding rate due to  thermal fluctuation. 


For BEMetaD and PBMetaD simulations, 18 backbone dihedral angles are used as CVs.
18 replicas are used and each replica lasts for 120 ns.
Other parameters are the same as in {\bf Peptoid trimer}.

{\bf Protein structure refinement.} For protein structure refinement, the initial structure the targets R0974s1, R0986s1 and R1002-D2 are obtained from the refinement category of the CASP13 competition. 
Then, they are solvated in (5.5 nm)${}^3$, (5.8 nm)${}^3$, (5.4 nm)${}^3$ dodecahedron boxes with 3384, 4191, 3334 water molecules, respectively.
Ions are add to neutralize the charge.
All simulations are carried out with the CHARMM36m force field~\cite{huang2017charmm36m} and the TIP3P water model~\cite{jorgensen1983comparison}.
The temperature is maintained at 300 K. 
136, 182 and 116 backbone dihedral angles are set as CVs.
8 walkers are used and the biased MD simulation lasted for 6 ns in each iteration.
The CV values are computed every 6 ps.
All other parameters are the same as in {\bf Peptoid} except that the widths of the hidden layers are chosen as (M1, M2, M3, M4) = (1200, 1200, 1200, 1200) for the  DNN models.
The flat-bottom harmonic restraint potential applied to each C$\alpha$ atom is defined by
\begin{linenomath*}
 \begin{eqnarray}\label{eqn:harm-pot}
 V_{fb}(\bm{r}_i)=\frac{1}{2}k_{fb}[d_g(\bm{r}_i;\bm{R}_i)-r_{fb}]^2H[d_g(\bm{r}_i;\bm{R}_i)-r_{fb}],
\end{eqnarray}
\end{linenomath*}
where $\bm{R}_i$ denotes the reference position of the $i$-th C$\alpha$ atom,  $r_{fb}$ is the distance from the center with a flat potential, $k_{fb}$ denotes the force constant,  $H$ denotes the Heaviside  function, and $d_g(\bm{r}_i;\bm{R}_i)$ is the distance from the reference position.
Here we set $r_{fb}$ to 0.6 nm and  $k_{fb}$ to 12 kJ/mol/nm$^2$.

{\bf MCMC simulations.}
We notice that the FES is a logarithm of the probability distribution, i.e.
\begin{align}
    A(\bm s) = -k_b T\ln p(\bm s) 
\end{align}
thus the low-dimensional FES is computed from the marginal distribution of the high-dimensional probability distribution corresponding to the high-dimensional FES. 
In this work, we use the Markov chain Monte Carlo (MCMC) method to calculate the marginal distribution. 

For example,  we have an $M$-dimensional CV space $\bm s=(s_1, \dots, s_M)$, and want to calculate the FES $A(s_1)$ from $A(s_1, \dots, s_M)$.
Since we have the definition for the marginal distribution on $s_1$
\begin{align}\label{eqn:fes-p}
    p(s_1) = \int p(s_1, \dots, s_M) ds_2\dots ds_M
\end{align}
then the FES on $s_1$ is given by
\begin{align}\label{eqn:fes}
    A(s_1) = -k_bT \ln \int e ^{-\frac1{k_bT}A(s_1, \dots, s_M)} ds_2\dots ds_M.
\end{align}
The integration in \eqref{eqn:fes-p} or \eqref{eqn:fes} is computed by MCMC.
Here, we carried out 2000 independent MC samplers and each lasts $10^6$ steps.
The example can be easily generalized to any FES defined on a low-dimensional subspace of the high-dimensional CV space. 

Note that the MCMC sampling is performed on the FES, so it is very efficient. 
The convergence of the MCMC simulations is assessed in Supplementary~Figure~5.

\textbf{A guideline for tuning the hyper-parameters of the adaptive RiD.}
In the hyper-parameter tuning procedure, we first estimate the statistical error introduced by the restrained MD simulations. This error defines the highest accuracy achievable by our DNN representation. Then the hyper-parameters like the batch size, the start learning rate, and the learning rate decay speed, are tuned to minimize the number of epochs needed for achieving the optimal accuracy. 
In the first few adaptive RiD iterations, the width of the DNN may be chosen to be a relatively small value, for example, 100 each hidden layer. 
As the adaptive RiD goes on, more parts of the FES are explored and the training data accumulate, the accuracy of the DNN model is observed to decrease. 
This indicates that the current DNN architecture is not powerful enough compared with the complexity of the explored FES. 
We stop the adaptive RiD manually when the relative error reaches $\sim$0.5, enlarge the DNN with a sub-network initialized by the original DNN, and then retrain the new DNN. 
This can substantially reduce the error of DNN and the adaptive RiD can continue.   
It is noted that the strategy of gradually enlarging DNN helps to determine the size of DNN but is not necessary, because using a large enough DNN at the beginning would not cause any difficulty in training the DNN.

\section{Data Availability}
The initial files of all examples for running adaptive RiD are available 
at Zenodo.\cite{data}
Our peptoid models come from Weiser's work\cite{weiser2017molecular}. Our chignolin model is obtained from Protein Data
608 Bank (PDB ID: 5AWL). The MD trajectories of chignolin from Anton can be obtained from Kresten et al.\cite{lindorff2011fast}.
The targets in CASP13 can be obtained from the official CASP list (\url{https://predictioncenter.org/casp13/targetlist.cgi}). The Markov state models of R0974s1, R0986s1, and R1002-D2 can be obtained from the work of Heo et al.\cite{heo2019driven}.
The input PLUMED2 files are available via  Plumed Nest with plumID:21.034. Source Data for Figures 2-6 is available with this manuscript.

\section{Code Availability}
Python implementation of our codes are available at github (\url{https://github.com/dongdawn/rid}) and Zenodo\cite{code}.


\end{document}